\documentstyle[12pt,aps]{revtex}

\begin{document}
\draft
\title{An illustration of chiral fermions on a 1+1 dimensional lattice}
\author{She-Sheng Xue
}
\address{
ICRA, INFN  and
Physics Department, University of Rome ``La Sapienza", 00185 Rome, Italy
}


\maketitle

\centerline{xue@icra.it}

\begin{abstract}

The vectorlike doubling of low-energy excitations is in fact a natural consequence of the pair-production around the zero-energy ($E=0$) due to the quantum field fluctuations of the lattice regularized vacuum. On the 1+1 dimensional lattice, we study an anomaly-free chiral model (11112) of four left-movers and one right-mover with strong interactions. Exact computations of relevant $S$-matrices illustrate that for high-momentum states, a negative energy-gap ($E<0$) develops; the bound state and its constituents, which have the same quantum numbers but opposite chiralities, fill the same energy-state so that chiral symmetries are preserved; for low-momentum states, the negative energy-gap vanishes and the bound state dissolves into its constituents near zero energy. As a consequence of the gauge-anomaly cancellation and the index theorem for flavor-singlet anomalies, the net number of zero-modes pushed down into and pumped out from the zero-energy level by the gauge field is zero.

\end{abstract}


\narrowtext

The parity-violating feature in the low-energy is strongly phenomenologically supported. Based on this feature, the successful standard model for particle physics is constructed in the form of a renormalizable quantum field theory with chiral gauge symmetries on the space-time. While, the very-small-scale structure of the space-time, the arena of physical reality, can exhibit rather complex structure of a space-time ``string" or ``foam", instead of a simple space-time point. As the consequence of these fundamental constituents of the space-time, the physical space-time gets endowed with a fundamental length and fundamental theory must be {\it finite}. However, with very generic axioms, the ``no-go" theorem \cite{nn81} demonstrates that the quantum field theories with parity-violating gauge symmetries, as the standard model, cannot be consistently regularized on the lattice, i.e., the discrete space-time. This paradox may imply a new physics beyond the standard model. 
Searching for a chiral gauge symmetric approach to properly regularize the standard model on the lattice has been greatly challenging to particle physicists for the last two decades. One of the classes of approaches is the modeling 
by appropriately introducing local interactions\cite{ep}-\cite{xue00}. However, the phenomenon of spontaneous symmetry breakings in the intermediate coupling and the argument of anomaly-cancellation within vectorlike spectra in the strong coupling prevent such modelings from a scaling region for low-energy chiral gauged fermions. It was then a general belief that it seems impossible to formulate a theory of exact chiral gauge symmetries on the lattice. Nevertheless, in refs.\cite{xue97}, a model with peculiar interactions and a plausible scaling region were advocated and the dynamics of realizing chiral gauged fermions was studied\cite{xue97l,xue00}. In this paper, based on a simple chiral model (11112) on the 1+1 dimensional lattice, we attempt to give an exact illustration of the dynamics realizing chiral gauge theories in the low-energy scaling region.

The chiral model (11112) is made of the $U(1)$ gauge field, four left-movers $\psi^i_L$ with charge $Q_L^i=1$ ($i=1,2,3,4)$ and one right-mover $\psi_R$ with charge $Q_R=2$. The t'Hooft condition for gauge anomaly cancellation $\sum_i(Q^i_L)^2=Q_R^2$ is satisfied.  With the fixed spatial and temporal lattice spacings $a$ and $a_t$ $(a\gg a_t)$, the free Hamiltonian is given by (we henceforth omit the index $i=1,2,3,4$),  
\begin{equation}
H_\circ ={1\over 2a}\sum_x\left(\bar\psi_L(x) D^L\cdot\gamma\psi_L(x)+\bar\psi_R(x) D^R\cdot\gamma\psi_R(x)\right),
\label{free}
\end{equation}
where all fermionic fields are two-component and dimensionless Weyl
fields, $x$ is the integer label of space sites, $\gamma$-matrix ($\gamma^2=1$) and  $D^{L,R}$ are ($\delta_{x,x\pm 1}\psi_L(x)=\psi_L(x\pm 1)$),
\begin{equation} 
D^{L,R}=([U_1(x)]^{Q_{L,R}}\delta_{x,x+1}
-[U_1^\dagger(x)]^{Q_{L,R}}\delta_{x,x-1}),
\label{kinetic} 
\end{equation} 
where $U_1(x)$ is the gauge field at a spatial link and the temporal gauge fixing $U_\circ(x)=1$. This is a chiral gauge model that cannot be naively quantized on the lattice due to the doubling problem.

The vectorlike doubling phenomenon is in fact the pair-production around zero energy ($E=0$) due to quantum field fluctuations of the lattice regularized vacuum (lattice vacuum) of eq.(\ref{free}). Since the 1+1 dimensional space-time is discretized with the lattice spacings $a$ and $a_t$, the volume of the lattice vacuum in the energy-momentum space ($E-\tilde p$ plan), $0\ge a_tE\ge-\pi$ and $|a\tilde p|\le\pi$ ($\hbar=1$), must be {\it finite}. The total number of negative energy states  must be {\it finite} for the reason that each state occupies a quantum volume of $h$. Each negative energy state is filled by both particle and antiparticle states so that the lattice vacuum exactly preserve chiral symmetries of (\ref{free}). As an example, the low-energy excitation of the right-mover ($\psi_R$) and its antiparticle can be created from just below zero energy to just above by quantum field fluctuations of the lattice vacuum, since these modes are massless. The energy and momentum fluctuations $\Delta E$ and $\Delta \tilde p$ of the lattice vacuum in such pair production process obey the Heisenberg uncertainty principle 
\begin{equation} 
a_t|\Delta E|\le \pi,\hskip0.5cm a|\Delta \tilde p|\le \pi.
\label{uncer} 
\end{equation}
These energy and momentum fluctuations $|\Delta E|,|\Delta \tilde p|$ of the lattice vacuum are apparently equal to the energy and momentum differences between particle and antiparticle created. 
Assuming the right-mover ($\psi_R$) with the energy $E=+\tilde p\simeq 0$ at the momentum state $a \tilde p\simeq 0$, its antiparticle (the left-mover) must be at the energy-momentum state $E= -\tilde p\simeq -\pi/a$ and  $a \tilde p\simeq +\pi$ according to eq.(\ref{uncer}).
However, the energy state $E= - \pi/a$ is actually an empty state (hole) of the lattice vacuum. All fully filled negative energy states with $E\in(0,-\pi/a)$ must fluctuate down to fill the empty states of lower negative energy levels, as a result the empty state moves to $E\sim 0$, that is just a low-energy excitation (doubler) of the antiparticle (left-mover) at $a \tilde p\simeq \pi$ and $E\sim 0$. Chiral symmetries are preserved for all negative energy states and the zero-energy state filled by pairs of left- and right- movers. All these discussions are applicable to four dimensions.  The total number of negative energy states of the vacuum is {\it infinite} in the continuum limit $a_t\rightarrow 0, a\rightarrow 0$ and the low-energy excitation (doubler) of the left-mover appear at $p\rightarrow\infty$. 

We introduce a neutral and massless spectator $\chi=\chi_L+\chi_R$, $\chi_R$ couples to four left-movers $\psi^i_L$ and $\chi_L$ couples to the right-mover $\psi_R$ as follow,
\begin{eqnarray}
H^L_i&=&g\sum_x \bar\psi_L(x)\cdot\left[\Delta\chi_R(x)\right]
\left[\Delta\bar\chi_R(x)\right]\cdot\psi_L(x),
\label{hil}\\
H^R_i&=&g\sum_x \bar\psi_R(x)\cdot\left[\Delta\chi_L(x)\right]
\left[\Delta\bar\chi_L(x)\right]\cdot\psi_R(x),
\label{hir}
\end{eqnarray}
where the multifermion coupling $g$ has dimension
$[a^{-1}]$ and the operator $\Delta$ is given as,
\begin{eqnarray}
\Delta\chi_{L,R}(x)&\equiv&
\left[ \chi_{L,R}(x+1)+\chi_{L,R}(x-1)-2\chi_{L,R}(x)\right],\nonumber\\
w(p)&=&{1\over2}\sum_xe^{-ipx}\Delta(x)=\left(\cos(p)-1\right),
\label{wisf}
\end{eqnarray}
where the dimensionless momentum $p=\tilde pa$. Eq.(\ref{wisf}) indicates that large momentum states of $\chi_R(\chi_L)$ strongly couple to $\psi_L(\psi_R)$, while small momentum states of $\chi_R(\chi_L)$ weakly couple to $\chi_R(\chi_L)$.
For the convenience of computations, we rescale the fermion fields $\psi\rightarrow (a_tg)^{1\over4}\psi$ and rewrite,
\begin{eqnarray}
a_tH={1\over 2a}({a_t\over g})^{1\over2}\sum_x\left(\bar\psi_L(x) D^L\gamma\psi_L(x)+\bar\psi_R(x) D^R\gamma\psi_R(x)
+\cdot\cdot\cdot\right)+a_tH^L_i+a_tH^R_i,
\label{action}
\end{eqnarray}
where $g$ in $H^{L,R}_i$ is rescaled away and ``$\cdot\cdot\cdot$" stands for the kinetic terms for $\chi$. We consider the limit $a_t/a\rightarrow 0$, $ga\rightarrow\infty$ and $ga_t$ is fixed.

This Hamiltonian system (\ref{action}) possesses the continuous Abelian chiral gauge symmetry, global chiral symmetries $U_{L,R}(1)$ and the shift-symmetries of $\chi_R$ and $\chi_L$\cite{gp}. Due to the Mermin and Wagner theorem\cite{nospon}, these continuous symmetries cannot be spontaneously broken for any values of the coupling $ga$. In addition, the shift-symmetries protect the right-mover $\psi_R$ sector and left-movers $\psi_L$ sector from coupling each other and guarantee the spectators $\chi_R, \chi_L$ decoupled as free particles\cite{xue97,gp}. These features greatly simply our illustrations. 

In this paper, we take the left-moving sector $\psi_L(x)$ as an example and analyze only the charged spectrum of Hamiltonian. For the strong coupling $ga\gg 1$, the three-fermion states $\Psi_R$ with the same quantum numbers of $\psi_L(x)$ is formed,
\begin{equation}
\Psi_R={1\over
2}(\bar\chi_R\cdot\psi_L)\chi_R,
\label{bound}
\end{equation}
which is a two-component Wely fermion state. To show this, we compute the S-matrix for this three-fermion state $\Psi_R(x,t)$,
\begin{equation}
S_{33}(x)\equiv \lim_{t_{f,i}\rightarrow \pm\infty}\langle\Psi_R(0,t_f)|\Psi_R(x,t_i)\rangle,
\label{sbound}
\end{equation}
where
\begin{equation}
|\Psi_R(x,t_i)\rangle=e^{iH_i^Lt_i}|\Psi_R(x,-\infty)\rangle;\hskip0.3cm
\langle\Psi_R(0,t_f)|=\langle\Psi_R(0,+\infty)|e^{-iH_i^Lt_f},
\label{asystate}
\end{equation}
and $|\Psi_R(x,\pm\infty)\rangle$ are the asymptotical states. We have then
\begin{equation}
S_{33}(x) = \lim_{t_{f,i}\rightarrow \pm\infty}\langle\Psi_R(0,+\infty)|e^{-iH_i^Lt_f}e^{iH_i^Lt_i}|\Psi_R(x,-\infty)\rangle.
\label{ssbound}
\end{equation}
Using ($\psi$ indicates $\psi_L$ or $\chi_R$)
\begin{equation}
\int d\psi(x,t)d\bar\psi(x,t)\bar\psi(x,t)\psi(x,t)\equiv |\psi(x,t)\rangle\langle \psi(x,t)|=1,
\label{com}
\end{equation}
we have
\begin{equation}
S_{33}(x) = \lim_{t_{f,i}\rightarrow \pm\infty}\langle\Psi_R(0,+\infty)|e^{-iH_i^Lt_f}|\psi(0,t_f)\rangle\langle \psi(0,t_f)|\psi(x,t_i)\rangle\langle \psi(x,t_i)|e^{iH_i^Lt_i}|\Psi_R(x,-\infty)\rangle.
\label{ssbound}
\end{equation}
We define the form factor of the three-fermion state $\Psi_R(x)$:
\begin{eqnarray}
Z_{t_i}(x)&\equiv&\langle \psi(x,t_i)|e^{iH_i^Lt_i}|\Psi_R(x,-\infty)\rangle
=\Pi_{-\infty}^{t_i}\int d\psi(x,t)d\bar\psi(x,t) e^{-a_tH_i^L}\bar\psi_L(x,t_i)\Psi_R(x,-\infty)\nonumber\\
Z_{t_f}(0)&\equiv&\langle\Psi_R(0,+\infty)|e^{-iH_i^Lt_f}|\psi(0,t_f)\rangle
=\Pi^{+\infty}_{t_f}\int d\psi(0,t)d\bar\psi(0,t) e^{-a_tH_i^L}\bar\Psi_R(0,+\infty)\psi_L(0,t_f),
\label{zz}
\end{eqnarray}
where we make the Wick rotation to the Euclidean space. The transfer matrix $S_{11}(x)$ of two intermediate states $|\psi(0,t_f)\rangle, |\psi(x,t_i)\rangle$ is given by
\begin{equation}
S_{11}(x)\equiv\langle \psi(0,t_f)|\psi(x,t_i)\rangle=\Pi_{t_f,x}^{t_i,y}\int d\psi(x,t)d\bar\psi(x,t) e^{-a_tH}\bar\psi_L(0,t_i)\psi_L(x,t_f),
\label{tran}
\end{equation}
where $t_{i,f}$ are finite and the $H$ is the total Hamiltonian.
As a result, the S-matrix can be written as,
\begin{equation}
S_{33}(x)= \lim_{t_{f,i}\rightarrow \pm\infty} Z_{t_i}(x)S_{11}(x)Z_{t_f}(0).
\label{tran}
\end{equation}

By eq.(\ref{zz}) we compute the form factors $Z_{t_i}(0)$, $Z_{t_f}(x)$: 
\begin{equation}
Z_{t_i}(0)=a_tg\Delta^2(0)\hskip0.5cm Z_{t_f}(x)=a_tg\Delta^2(x).
\label{rzz}
\end{equation}
To compute $S_{11}(x)$, we define another transfer matrix $S_{31}(x)$ of two intermediate states $|\Psi_R(x,t_i)\rangle, |\psi_L(0,t_f)\rangle $,
\begin{equation}
S_{31}(x)\equiv\langle\psi_L(0,t_f)|\Psi_R(x,t_i)\rangle.
\label{s31}
\end{equation}
The exact recursion relations for $S_{11}(x)$ and $S_{31}(x)$ can be obtained (cf. eqs.(C15), (C17) and (C18) in ref.\cite{xue97}) 
\begin{eqnarray} 
S_{11}(x)&=&{1\over a_tg\Delta^2(x)}\left({a_t\over 2a
}\right)^3\sum^\dagger S_{31}(x)\gamma,\label{re11}\\
S_{31}(x)&=&{1\over2}\left({\delta(x)\over 2a_tg\Delta^2(x)}
+{1\over a_tg\Delta^2(x)}\left({a_t\over 2a
}\right)\sum^\dagger S_{11}(x)\gamma\right),
\label{re21}\\
\end{eqnarray}
where $\sum^\dagger f(x)=f(x+1)-f(x-1)$. These recursion equations can be solved by the Fourier transformations for $p\not=0$,
\begin{eqnarray}
S_{11}(p)&=&\sum_x e^{-ipx}S_{11}(x)
={{ia_t\over 2a}\sin (p)\gamma\over
({a_t\over a})^2\sin^2(p)+M^2(p)},\label{sll21}\\
S_{31}(p)&=&\sum_xe^{-ipx}T_{31}(x) ={{1\over2}M(p)\over
({a_t\over a})^2\sin^2(p)+M^2(p)},\label{slm21}\\
M(p)&=&8agw^2(p).\label{m}
\end{eqnarray}

As a result, for $t_{f,i}\rightarrow \pm\infty$ we obtain the exact S-matrix (\ref{ssbound})
\begin{equation}
S_{33}(p)
=Z(p){{ia_t\over 2a}\sin (p)\gamma\over
({a_t\over a})^2\sin^2(p)+M^2(p)}Z(p),
\label{rsc11}
\end{equation}
where the form factor (\ref{rzz}) $Z(p)=4a_tgw^2(p)$ in the momentum space. The three-fermion state $\Psi_R$ is represented by the pole and its residual(form factor) in the S-matrix (\ref{rsc11}). We make a wave-function renormalization of 
three-fermion states $\Psi_R$ with respect to the doubler $p=\pi$,
\begin{equation}
\Psi_R|_{ren}=Z^{-1}(\pi)\Psi_R;\hskip1cm Z(\pi)=16a_tg.
\label{rbound}
\end{equation}
$\Psi_R|_{ren}$ mixes with $\psi_L$ to form a 
four-component massive Dirac fermion $\Psi_c$, 
\begin{equation}
\Psi_c=(\Psi_R|_{ren},\psi_L),
\label{di}
\end{equation}
represented by the pole of the S-matrix ,
\begin{eqnarray}
S_c(x)&=&\lim_{t_{f,i}\rightarrow \pm\infty}\langle\Psi_c(0,t_f)|\bar\Psi_c(x,t_i)\rangle\nonumber\\
S_c(p)&=&{{ia_t\over a}\sin (p)\gamma +M(p)\over
({a_t\over a})^2\sin^2(p)+M^2(p)}={1\over
-{ia_t\over a}\sin (p)\gamma +M(p)}
\label{ds}
\end{eqnarray}
at $p=\pi$ and mass ${8ag/ a_t}$.

The Hamiltonian $H_{\rm bound}$ in the basis of the eigen-states of $\psi_L$ and three-fermion state $\Psi_R$, i.e., Dirac fermion $\Psi_c$, can be obtained from eq.(\ref{ds}):
\begin{equation}
-a_tH_{\rm bound}=-{ia_t\over a}\sin (p)\gamma +M(p),\hskip0.3cm H_{\rm bound}={i\over a}\sin (p)\gamma -{M(p)\over a_t}.
\label{eff}
\end{equation}
This clearly indicates that the binding energy of the three-fermion state $\Psi_R$ is ${-M(p)/a_t}$.
Due to the locality of the
theory, the vector-like spectrum (\ref{eff})
obtained by the strong coupling for large momentum states, can be analytically
continued to small momentum states. However, eq.(\ref{ds}) does not represent a massless pole at $p=0$, since the S-matrix $S_{33}(p)$ (\ref{rsc11}) is no longer singular at $p=0$ for the form factor $Z(p)$ positively vanishing $O(p^4)$ and we are not allowed to make wave-function renormalization (\ref{rbound})\cite{note}. 

$\Psi_R(p)$ consists of three constituents $\psi_L(-p')$, $\chi_R(q)$ and $\bar\chi_R(-q')$ with the total momentum $p=-p'+q-q'$. $\bar\Psi_R(-p)$ consists of three constituents $\bar\psi_L(p')$, $\chi_R(q)$ and $\bar\chi_R(-q')$ with the total momentum $-p=p'+q-q'$.
The relative momentum $|\Delta p|$ of three constituents within the bound state $\Psi_R$ is of $O(|p|)$. According to the Heisenberg uncertainty principle $|\Delta p||\Delta x|\sim 2\pi a$, where $|\Delta x|$ is the size of the bound state $\Psi_R$. If $|\Delta p| \sim\pi$, the size $|\Delta x|$ of the bound state is a few of the lattice spacing $a$. As  
``$|p|$" goes to zero, the size $|\Delta x|$ of $\Psi_R$ increase, the negative energy-gap ${-M(p)/a_t}$ and form factor $Z(p)$ in (\ref{rsc11}) go to zero, the bound state $\Psi_R$ dissolves into its three constituents.
     
To illustrate this, we first define the notion of the three-fermion cut ${\cal C}[\Psi_R]$\cite{xue97l}, which is a virtual state rather than a particle state. For a given total momentum ``$\tilde p$" in the low-energy region, this virtual state ${\cal C}[\Psi_R](\tilde p)$ has the same total momentum ``$\tilde p$" and contains the same constituents as $\Psi_R(\tilde p)$: 
$\psi_L(-\tilde p')$, 
$\chi_R(\tilde q)$ and $\bar\chi_R(\tilde q')$, where $\tilde p=\tilde q-\tilde q'-\tilde p'$. $\psi_L(-\tilde p')$, $\chi_R(\tilde q)$ and $\bar\chi_R(\tilde q')$ are low-energy excitations.

This virtual state ${\cal C}[\Psi_R]$ has the same quantum numbers as the bound state $\Psi_R$.
The total energy of such a virtual state ${\cal C}[\Psi_R]$ is given by
\begin{eqnarray}
E_t&=&E_1(\tilde q)+E_2(\tilde q')+E_3(\tilde p'),\nonumber\\
E_1(\tilde q)&=& \tilde q>0,\hskip0.2cm \tilde q>0 \hskip0.2cm {\rm for} \hskip0.2cm \chi_R, \nonumber\\ 
E_2(\tilde q')&=& -\tilde q'>0,\hskip0.2cm \tilde q'<0 \hskip0.2cm {\rm for} \hskip0.2cm \bar\chi_R, \nonumber\\ 
E_3(\tilde p')&=&-\tilde p'>0,\hskip0.2cm \tilde p'<0 \hskip0.2cm {\rm for} \hskip0.2cm \psi_L. 
\label{totale}
\end{eqnarray}
 There is no any definite one to one 
relationship between the total energy $E_t$ and the total momentum ``$\tilde p$". The total energy spectrum $E_t$ of such a virtual state ${\cal C}[\Psi_R]$ is continuum with respect to the given total momentum ``$\tilde p$".  
The lowest energy $E^{\rm min}_t(\tilde p)$ (the energy-threshold) of the virtual state ${\cal C}[\Psi_R]$
is
\begin{equation}
E^{\rm min}_t(\tilde p)=|\tilde p|\le E_t= |\tilde p'|+|\tilde q|+|\tilde q'|.
\label{thre} 
\end{equation}

Given the same total momentum ``$p$" in the low-energy, the bound state $\Psi_R(\tilde p)$ is stable, only if only there is an energy gap $\Delta(\tilde p)$ between the energy-threshold (\ref{thre}) of ${\cal C}[\Psi_R](\tilde p)$ and the negative binding-energy (\ref{eff})
of $\Psi_R(\tilde p)$, i.e., 
\begin{equation}
\Delta(\tilde p)= E^{\rm min}_t(\tilde p)-(-{M(\tilde p)\over a_t})>0.
\label{sta}
\end{equation}
As the energy gap $\Delta(\tilde p)$ vanishes for $a\tilde p\rightarrow 0$, the three-fermion state $\Psi_R(\tilde p)$ must dissolve into its virtual state ${\cal C}[\Psi_R](\tilde p)$. $\Delta(\tilde p)=0$ determining the critical momentum threshold $|\tilde p_c|$ for such a dissolving phenomenon in the low-energy limit, 
\begin{equation}
|\tilde p_c|={1\over a}\left({a_t\over 2a^2g}\right)^{1\over3}\ll {\pi\over2a},\hskip0.3cm ag\gg 1, \hskip0.3cm{a_t\over a}\ll 1,
\label{con}
\end{equation}
and the energy-threshold $\epsilon=\sqrt{2}|\tilde p_c|$. As an example, for $ag=100$ and $a_t/a=10^{-1}$, $\epsilon\simeq
10^{-1}a$.
This dissolving phenomenon is chiral symmetric, since $\Psi_R(\tilde p)$ and ${\cal C}[\Psi_R](\tilde p)$ have the same quantum numbers.

Let us discuss how the lattice vacuum is filled by the large momentum states of $\Psi_R$. For simplicity, we introduce the notation $\psi(p)$ indicating fermion $\psi$ at the momentum state ``$p$" ($\pi>p>0$). As $\psi_L(-p)$ mixes up with $\Psi_R(p)$ to form the Dirac fermion $\Psi_c(p)$ (\ref{di}), $\psi_L(-p)$ and $\Psi_R(p)$ fill into the same positive energy state. $\bar\psi_L(p)$ and $\bar\Psi_R(-p)$ fill into the same negative energy state. Thus, each energy state is filled by both left- and right- moving states with the same quantum numbers so that chiral symmetries is preserved by the vectorlike spectrum $(\psi_L(-p),\Psi_R(p))$. In addition, a negative energy-gap ${-M(p)/a_t}$ is formed.
The gauge potential $A_\circ=0$ and $A_1$ give the electric field ${\cal E}=\partial_\circ A_1$ in the direction of $p>0$. The electric force $\pm Q{\cal E}$, sign ``$+(-)$" for the right(left)-mover, is the rate of changing momentum states $``p"$ by the unit of $2\pi$ per unit volume of space-time\cite{nn91}. The electric force drives fermions spectra of high-momentum states $``p"$ flowing along their dispersion relations. In $ p\le -a|\tilde p_c|$, the electric field ${\cal E}$ pushes the four left-movers $\psi^i_L(p)$ down to the energy-threshold $\epsilon$ and pumps the four three-fermion states $\bar\Psi^i_R(p)$ up to the energy-threshold $-\epsilon$. In $ p\ge a|\tilde p_c|$, the electric field ${\cal E}$ pumps the four three-fermion states $\Psi^i_R(p)$ away from the energy-threshold $\epsilon$ and pushes the four left-movers $\bar\psi^i_L(p)$ down below the energy-threshold $-\epsilon$.
The rate of pumping out four three-fermion states $\bar\Psi^i_R(p)$ in $ p<-a|\tilde p_c| $ and pushing down four left-movers $\bar\psi^i_L(p)$ for $ p>a|\tilde p_c| $ are the exactly same, consistently with the finite number of states of the lattice vacuum.   

The discussions and computations for the right-mover $\psi_R$, the corresponding three-fermion state $\Psi_L={1\over
2}(\bar\chi_L\cdot\psi_R)\chi_L$ and cut ${\cal C}[\Psi_L]$ are the exactly same as that for the left-moving sector $\psi_L$. 
 
Now we focus the spectral flows of low-energy excitations within $-\epsilon<\tilde p<\epsilon$. The neutral spectator fermion $(\chi_L,\chi_R)$ is a free and massless Dirac particle ($E=\pm \tilde p$). Since the vacuum energy states including the zero-energy state ($E\in [0,-\epsilon)$) are only filled by the left(right)-movers $\psi^i_L(-\tilde p)(\psi_R(\tilde p))$ without their partners of the same charge and opposite chirality, the $U(1)$ chiral gauge symmetry is broken and gauge anomalies must appear. In the gauge fixing $A_\circ=0$, the gauge anomalies are given by  
$-(Q^i_L)^2{\cal E}/(4\pi)$ for the gauge current $J^i_L=Q_L^i\bar\psi^i_L\gamma\psi^i_L$ of four left-movers $\psi^i_L$ and $+ (Q_R)^2{\cal E}/(4\pi)$ for the gauge current $J_R=Q_R\bar\psi_R\gamma\psi_R$ of right-mover $\psi_R$, which are proportional to the rate of pushing the charges $Q_L^i$ of four $\psi^i_L$ down into and pumping the charge $Q_R$ of one $\psi_R$ out from the zero-energy level $E=0$. Due to the fact that the t'Hooft condition is obeyed and gauge anomalies are exactly canceled, the corresponding net charges pushed down into and pumped out from the lattice vacuum by the electric field ${\cal E}$ are zero. However, the corresponding net number of zero modes pushed down into and pumped out from the lattice vacuum is not zero, i.e., 4-1=3. It seems to be inconsistent for the finiteness of the lattice vacuum for no extra rooms accommodating 3 zero modes. 

On the other hand, the vacuum energy states ($E\in [0,-\epsilon)$) that are filled by the left(right)-movers $\psi^i_L(-\tilde p)(\psi_R(\tilde p))$ break the global chiral symmetries $U_{L,R}(1)$. The divergence of Noether currents of these global symmetries receives anomalies. Such flavor-singlet anomalies are given by $\pm {\cal E}/(4\pi)$, ``+" for $j_L=\sum_i\bar\psi^i_L\gamma\psi^i_L$ and ``-" for $j_R=\bar\psi_R\gamma\psi_R$. $\pm {\cal E}/(4\pi)$ are proportional to the rate of pushing the number of state of left-movers down into and pumping the number of state of right-mover out from  zero energy $E=0$. By the index theorem, the axial anomaly of the current $j^5=j_R-j_L$ is given by $\Delta n=n_--n_+$, where $n_-(n_+)$ is the number of right(left)-movers. $\Delta n$ are the fermion numbers carried by topological gauge fields. $\Delta n=1-4=-3$ indicating three zero modes flowing out from the lattice vacuum and three states are emptied for accommodating 3 zero modes pushed in, consistently with the finiteness of the lattice vacuum. 

In fact, for the reason that the dissolving energy-scale $\epsilon\ll\pi/a\ll\pi/a_t$ and the asymmetry $a\gg a_t$ in the space-time turns out to be irrelevant at the low-energy scale $\epsilon$,
we can define a low-energy effective Lagrangian for the 11112 model with a continuous regularization at the energy scale $\epsilon$.
The asymmetry of filling the vacuum energy states ($E\in [0,-\epsilon)$), as discussed in above, must appear as explicit symmetry breaking terms in the low-energy effective Lagrangian at the scale $\epsilon$. As results we have, (i) the gauge anomalies $ Q_L^i\epsilon^{\mu\nu}\partial_\mu A_\nu/(4\pi)$ and $Q_R\epsilon^{\mu\nu}\partial_\mu A_\nu/(4\pi)$ of the $U(1)$ gauge symmetry; (ii) the flavor-singlet anomalies $\pm\epsilon^{\mu\nu}F_{\mu\nu}/(4\pi)$ of the $U_{L,R}(1)$ global chiral symmetries and the axial anomaly is given by  $\Delta n=n_--n_+$; (iii) local counterterms at the scale $\epsilon$.



\begin{references}

\bibitem{nn81}
H.B.~Nielsen and M.~Ninomiya, Nucl.~Phys.~B185 (1981) 20, {\it
ibid.} {\bf B193} (1981) 173, Phys.~Lett. B105 (1981) 219.

\bibitem{ep}
E.~Eichten and J.~Preskill, Nucl.~ Phys. B268 (1986) 179.\\
J.~Smit, Acta Physica Polonica B17 (1986) 531;\\
P.D.V.~Swift, Phys.~Lett. B145 (1984) 256;\\
S.~Aoki, I-Hsiu Lee and S.-S.~Xue, Phys.~Lett. B229 (1989) 79
and BNL Report 42494 (1989).\\
G.~Preparata and S.-S.~Xue, Phys.~Lett. B264 (1991) 35.\\
I.~Montvay, Phys.~Lett. B199
(1987) 282; Nucl. Phys. B29 (Proc.~Suppl.)
(1992) 159, {\it ibid} 30B (1993) 621 
and references therein.\\
M.F.L.~Golterman, D.N.~Petcher and E.~Rivas, Nucl.\ Phys. B395 
(1993) 597.\\
D.N.~Petcher, Nucl.~Phys. (Proc.~Suppl.) B30 
(1993) 52, references there in.\\
M.~Creutz, C.~Rebbi, M.~Tytgat and S.-S.~Xue,
Phys.~Lett. B402 (1997) 341.

\bibitem{xue97}
S.-S.~Xue, Phys.~Lett. B381 (1996) 277 and Nucl.~Phys. B486 (1997) 282.

\bibitem{xue97l}
S.-S.~Xue, Phys.~Lett. B408 (1997) 299; 
Nucl.~Phys. (Proc.~Suppl.) B53 (1997) 668.

\bibitem{xue00}
S.-S.~Xue, Phys.~Rev. D61 (2000) 054502, Nucl.~Phys. B486 (2000) 282 and references there in.

\bibitem{gp}
M.F.L.~Golterman, D.N.~Petcher,  Phys.~Lett. B225 
(1989) 159.

\bibitem{nospon}
N.D.~Mermin and H.~Wagner, Phys.~Rev.~Lett.~17 (1966) 113;\\
S.~Coleman, Comm.~Math.~Phys., 31 (1973) 259.

\bibitem{note}
It is worthwhile to mention that if the operator $\Delta=1$ in the interacting Hamiltonian $H_i^L$ in eq.(\ref{hil}), the same computations lead to the bound states, e.g., $\Psi_R(x)$ with the binding energy $-2ag/a_t$ and the form factor $Z=a_tg$ at $p=0$, indicating a massive Dirac fermion at $p=0$.

\bibitem{nn91}
H.B.~Nielsen and M.~Ninomiya, Int.~J.~ of Mod.~Phys. A6 (1991) 2913.



\end{references}
\end{document}